\documentclass[aps,prd,superscriptaddress,showpacs,nofootinbib,amssymb,amsfonts,amssymb,amsmath,twocolumn,balancelastpage,floatfix]{revtex4-1}

\usepackage{amssymb}
\usepackage{lipsum}
\usepackage{microtype}
\usepackage{graphicx}
\usepackage{float}
\usepackage{wrapfig}
\usepackage{bm}
\usepackage{color}
\usepackage{comment}
\usepackage{xcolor}
\usepackage[normalem]{ulem}
\usepackage[colorlinks=true,urlcolor=blue,linkcolor=blue,citecolor=blue,hypertexnames]{hyperref}
\usepackage{mathrsfs}
\usepackage[version=3]{mhchem}
\usepackage{booktabs}
\usepackage{natbib}
\usepackage{dcolumn}% Align table columns on decimal point
\usepackage{url}
\usepackage[nameinlink]{cleveref}
\hypersetup{
colorlinks=true,
citecolor=blue,
citebordercolor=red,
linktoc=all,
linkcolor=blue,
urlcolor=blue
}
\newcommand{\Slash}[1]{{\ooalign{\hfil/\hfil\crcr$#1$}}} 

\newcommand{\p}{\partial}

\newcommand{\nn}{\nonumber\\}

\newcommand{\df}{\text{d}}

\newcommand{\tr}{{\rm tr}\,}

\newbox{\ORCIDicon}
\sbox{\ORCIDicon}{\large\includegraphics[width=0.8em]{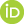}}

\begin{document}

\title{Emergence of dynamical tensor fields in composite models of gravity}

\author{Yadikaer \surname{Maitiniyazi}\,\href{https://orcid.org/0009-0004-0826-1130}{\usebox{\ORCIDicon}}}
\email{ydqem22@mails.jlu.edu.cn}
\affiliation{Center for Theoretical Physics and College of Physics, Jilin University, Changchun 130012, China}
\author{Masatoshi \surname{Yamada}\,\href{https://orcid.org/0000-0002-1013-8631}{\usebox{\ORCIDicon}}}
\email{m.yamada@kwansei.ac.jp}
\affiliation{Department of Physics and Astronomy, Kwansei Gakuin University,
Sanda, Hyogo, 669-1330, Japan}

\begin{abstract}
We investigate composite models of gravity and explore how dynamical tensor fields can emerge within the functional renormalization group framework. 
We consider two prototype models: a fermionic theory and a scalar theory. 
In both cases, an auxiliary tensor field is introduced via a Hubbard-Stratonovich transformation, corresponding to the composite channel associated with the energy-momentum tensor. 
We derive the flow equations for the field renormalization factors of the auxiliary tensor field and demonstrate that finite kinetic terms are dynamically generated in the infrared regime. 
Agreement with the diffeomorphism-invariant quadratic Einstein-Hilbert structure can be established in the transverse-traceless spin-2 sector, while the remaining contributions reside in the longitudinal and trace sectors. 
Although these terms can be cast into a form reminiscent of gauge-fixed presentations of the quadratic Einstein-Hilbert action, we do not interpret them as originating from a genuine gauge-fixing procedure within the present truncation.
\end{abstract}

\maketitle

%% main text%%%%%%%%%%%%%%%%%%%%%%%%%
\section{Introduction}
%%%%%%%%%%%%%%%%%%%%%%%%%%%%%%%%%%%%%

The nature of gravity remains one of the biggest mysteries in fundamental physics. While general relativity succeeds remarkably well in matching experimental observations, a corresponding quantum theory of gravity remains elusive.
The quantization of the metric field based on perturbation theory within the Einstein-Hilbert action leads to ultraviolet divergences which cannot be eliminated by renormalization.
Such an incompatibility with quantum mechanics has been motivating searches for a microscopic description of gravity beyond the framework of quantum field theory. 

Although it is clear that on macroscopic scales gravitational interactions are mediated by the metric field, there is a question of whether the metric field is a fundamental field at the microscopic level.
Among the various approaches to quantum gravity, composite models offer an intriguing possibility that the graviton is not an elementary particle, but rather a bound state or emergent excitation arising from more fundamental degrees of freedom.

The idea that gravity may be emergent is inspired by parallels with other known composite phenomena in physics. 
For instance, the pions in quantum chromodynamics and the Nambu--Jona-Lasinio model \cite{Nambu:1961tp,Nambu:1961fr} are low-energy excitations of composite fermions. 
These analogies have motivated the exploration of models in which the graviton arises from strong dynamics or large-$N$ gauge theories, or collective excitations in analog spacetime systems.
Such approaches can be viewed as possible realizations of pregeometry \cite{Terazawa:1976eq,Terazawa:1977xa,Akama:1977hr,Akama:1977nw,Akama:1978pg,Terazawa:1980vf,Akama:1981dk,Akama:1979tm,Akama:1981kq,Akama:1990ur,Terazawa:1991yj,Akama:1991pv,Akama:1991np,Akama:1994ehj,Wetterich:2003wr,Hebecker:2003iw,Wetterich:2004za,Wetterich:2021ywr,Wetterich:2021cyp,Wetterich:2021hru,Floreanini:1990cf,Matsuzaki:2020qzf,Volovik:2021wut,Maitiniyazi:2023hts,Maitiniyazi:2024zty,Oda:2025soa,Adami:2025pqr}.
Moreover, some models may also provide a natural explanation for the weakness of gravity compared to other interactions, as gravitational interactions could be suppressed by compositeness or large scales associated with binding.

In this paper we study the emergence of dynamical tensor fields in composite gravity models by utilizing the functional renormalization group (fRG). We begin by outlining general principles and constraints on graviton compositeness, followed by a detailed analysis of representative models.
We introduce auxiliary tensor fields corresponding to the energy-momentum tensor via the Hubbard-Stratonovich transformation. As a first step toward a consistent composite theory of gravity, we study how interactions with fundamental fields generate kinetic terms for these tensor fields through their interactions with the underlying dynamical fundamental fields.
We also compare the induced kinetic terms to those in general relativity.

This paper is organized as follows: In \Cref{sec:Models}, we introduce the composite models of gravity and present the setups in this work.
In \Cref{sec:FRGE}, we derive the flow equations for the mass parameters and the field renormalization factors of an auxiliary tensor field corresponding to the composite field.
We solve these flow equations and analyze the structure of the kinetic terms of the composite field in \Cref{sec:Analysis}.
Finally, \Cref{sec:summary} is devoted to summarizing the results.

%%%%%%%%%%%%%%%%%%%%%%%%%%%%%%%%%%%%
\section{Models}
\label{sec:Models}
%%%%%%%%%%%%%%%%%%%%%%%%%%%%%%%%%%%%
We study two models: a fermionic theory and a bosonic (scalar) theory.
In this section, we present their initial action and introduce an auxiliary tensor field via the Hubbard-Stratonovich transformation.

%%%%%%%%%%%%%%%%%%%%%%%%%%%%%%%%%%%%%
\subsection{Fermionic model}
%%%%%%%%%%%%%%%%%%%%%%%%%%%%%%%%%%%%%

The first case of composite models is based on a $U(N_f)$ fermionic theory in Euclidean space in which the metric is $\delta_{\rho\sigma}=(1,1,1,1)$.\footnote{We adopt the definition of Euclidean space as in \cite{Huang:2024ypj}.}
The starting action is given by
\begin{align}
S = \int\df^4x\left[\frac{1}{2}\bar\psi \overleftrightarrow{\Slash \p} \psi
- f_2 T^{(f)}_{\mu\nu}T^{(f)}_{\mu\nu} - f_0 \left(T^{(f)}_{\rho\rho} \right)^2
\right]\,,
\label{eq:StartingFerminAction}
\end{align}
where $\psi=(\psi_0,\ldots, \psi_{N_f})$ is the $N_f$-flavor spinor fields and $T^{(f)}_{\mu\nu}$ denotes the chosen symmetric spin-2 fermion bilinear composite operator (written in the standard free-field form),
\begin{subequations}
\begin{align}
&T^{(f)}_{\mu\nu} = \frac{1}{4}\bar\psi \left( \gamma_\mu \overleftrightarrow{\p_\nu} + \gamma_\nu \overleftrightarrow{\p_\mu} \right)\psi\,,\\
&T^{(f)}_{\rho\rho} = \delta_{\rho\sigma}T^{(f)}_{\rho\sigma}\,.
\label{eq: EM tensor for fermion}
\end{align}
\end{subequations}
Here, $\bar\psi \gamma_\nu\overleftrightarrow{\p_\mu} \psi=\bar\psi \gamma_\nu ({\p_\mu} \psi) - ({\p_\mu}\bar\psi)\gamma_\nu  \psi$ and $\Slash \p= \delta_{\mu\nu}\gamma_\mu \p_\nu=\gamma_\mu \p_\mu$ with $\gamma_\mu$ being the Dirac matrices.
Note that $f_2$ and $f_0$ have mass dimension $-4$.

Introducing an auxiliary tensor field $H_{\mu\nu}$ in the path integral formalism, the action \labelcref{eq:StartingFerminAction} can be rewritten as
\begin{align}
S = \int\df^4x\left[\frac{1}{2}\bar\psi \overleftrightarrow{\Slash \p} \psi + T^{(f)}_{\mu\nu} H_{\mu\nu}
- \lambda_2' H_{\mu\nu}H_{\mu\nu} - \lambda_0'H^2
\right]\,,
\label{eq:RewrittenFerminAction}
\end{align}
where $H=\delta_{\rho\sigma}H_{\rho\sigma}=H_{\rho\rho}$ is the trace mode in the tensor field $H_{\mu\nu}$.
Here, the tensor field $H_{\mu\nu}$ is dimensionless, and $\lambda_2'$ and $\lambda_0'$ have mass dimension $+4$.
Note that using the equation of motion for $H_{\mu\nu}$
\begin{align}
H_{\mu\nu} = \frac{1}{2\lambda_2'}\left( T^{(f)}_{\mu\nu} - \frac{\lambda_0'}{\lambda_2'+4\lambda_0'}\delta_{\mu\nu} T^{(f)}{}^\rho{}_\rho \right)\,,
\end{align}
the action \labelcref{eq:RewrittenFerminAction} returns to the original one \labelcref{eq:StartingFerminAction} and it turns out that the relations between couplings are given by $f_2=-1/(4\lambda_2')$ and $f_0=-\lambda'_0/(4\lambda'_2(\lambda'_2+4\lambda'_0))$.

%%%%%%%%%%%%%%%%%%%%%%%%%%%%%%%%%%%%%
\subsection{Scalar model}
%%%%%%%%%%%%%%%%%%%%%%%%%%%%%%%%%%%%%

The second case is the massless $O(N)$ scalar field whose action in Euclidean space is given by
\begin{align}
S = \int\df^4x\left[\frac{1}{2} (\p_\mu \vec\phi)^2 
- g_2 T^{(s)}_{\mu\nu}T^{(s)}_{\mu\nu} - g_0 \left( T^{(s)}_{\rho\rho} \right)^2
\right]\,,
\label{eq:StartingBosonAction}
\end{align}
where $\vec\phi=(\phi_1,\cdots,\phi_N)$ is an $N$-component scalar field, $(\p_\mu \vec\phi)^2 =\p_\mu \vec\phi\cdot \p_\mu \vec\phi$ and the symmetric spin-2 scalar bilinear composite operator $T^{(s)}_{\mu\nu}$ is given by
\begin{subequations}
\begin{align}
&T^{(s)}_{\mu\nu} =\p_\mu \vec \phi\cdot \p_\nu \vec \phi - \frac{\delta_{\mu\nu}}{2}(\p_\rho \vec \phi)^2 \,,\\
&T^{(s)}_{\rho\rho} = \delta_{\rho\sigma}T^{(s)}_{\rho\sigma}\,. 
\label{eq: EM tensor for scalar}
\end{align}
\end{subequations}
We introduce an auxiliary tensor field $C_{\mu\nu}$ and rewrite the action as
\begin{align}
S = \int\df^4x\left[\frac{1}{2} \p_\mu \vec\phi \cdot \p_\mu \vec\phi
+  T^{(s)}_{\mu\nu} C_{\mu\nu}
- \xi_2' C_{\mu\nu}C_{\mu\nu} - \xi_0' C^2
\right]\,,
\label{eq:RewrittenScalarAction}
\end{align}
where $C=\delta_{\rho\sigma}C_{\rho\sigma}=C_{\rho\rho}$.
In the same manner as in the fermionic case, the use of the equation of motion for $C_{\mu\nu}$,
\begin{align}
C_{\mu\nu} = \frac{1}{2\xi_2'}\left( T^{(s)}_{\mu\nu} - \frac{\xi_0'}{\xi_2'+4\xi_0'}\delta_{\mu\nu} T^{(s)}_{\rho\rho} \right) \,,
\end{align}
allows us to reproduce \labelcref{eq:StartingBosonAction} with the relations between couplings as $g_2=-1/(4\xi_2')$ and $g_0=-\xi_0'/(4\xi_2'(\xi_2'+4\xi_0'))$.
Note here that $g_2$ and $g_0$ are mass dimension $-4$. Accordingly, $\xi_2'$ and $\xi_0'$ are mass dimension $+4$, while $C_{\mu\nu}$ is a dimensionless field.

%%%%%%%%%%%%%%%%%%%%%%%%%%%%%%%%%%%%
\section{Functional flow equations}
\label{sec:FRGE}
%%%%%%%%%%%%%%%%%%%%%%%%%%%%%%%%%%%%
In the actions \labelcref{eq:RewrittenFerminAction,eq:RewrittenScalarAction}, the tensor fields, $H_{\mu\nu}$ and $C_{\mu\nu}$, are nondynamical fields, i.e., there are no kinetic terms.
Our aim in this work is to investigate the ``dynamicalization" of the tensor fields $H_{\mu\nu}$ and $C_{\mu\nu}$ thanks to quantum fluctuations of the fermionic field and scalar field in the models \labelcref{eq:RewrittenFerminAction,eq:RewrittenScalarAction}.
To this end, we utilize the Wetterich equation \cite{Wetterich:1992yh} which describes the deformation of the one-particle irreducible effective action $\Gamma_k$ defined at an infrared (IR) cutoff scale $k$. See also \cite{Ellwanger:1993mw, Morris:1993qb}.
Its form is given by
\begin{align}
\label{eq:Wettericheq}
\p_t \Gamma_k = \frac{1}{2}\text{STr}\left[ \frac{1}{\Gamma_k^{(2)} + \mathcal R_k}\p_t \mathcal R_k \right]\,.
\end{align}
Here $t=\ln (k/\Lambda)$ is the dimensionless scale with a reference scale $\Lambda$, $\Gamma_k^{(2)}$ is the full two-point function obtained by the second-order functional derivative  for $\Gamma_k$ with respect to the superfield $\Phi=(\psi,\,\bar\psi,\,\vec\phi, \,H_{\mu\nu},\, C_{\mu\nu},\cdots)$, and STr denotes the functional supertrace acting on all spaces (spinor, momentum, etc.) in which the superfield $\Phi$ is defined.
In the formulation \labelcref{eq:Wettericheq}, the coarse-graining process is realized by the regulator function $\mathcal R_k$ which is written as 
\begin{align}
\mathcal R_k(p) 
=
\begin{cases}
 i\Slash p\, r_k^\psi(|p|/k)& \text{for fermion}\,, \\[2ex]
 p^2 \, r_k^\phi(|p|/k) &  \text{for scalar}\,.
\end{cases}
\end{align}
In this work, for $r_k(|p|/k)$, we employ the Litim type regulator \cite{Litim:2001up} for the fermionic and scalar fields respectively as
\begin{align}
\label{eq:fermioncut}
r_k^\psi(|p|/k)&=\left( \frac{k}{|p|} -1 \right)\theta\left(1-|p|^2/k^2\right)\,, \\[2ex]
\label{eq:scalarcut}
r_k^\phi(|p|/k)&=\left( \frac{k^2}{|p|^2} -1 \right)\theta\left(1-|p|^2/k^2\right)\,,
\end{align}
where $\theta(x)$ is the step function
\begin{align}
    \theta(x) = \begin{cases}
        1, &  x \geq 0\,, \\[1ex]
        0, &  x < 0\,.
    \end{cases}
    \label{eq:Stepfunc}
\end{align}
In what follows, we present an ansatz for the effective action $\Gamma_k$ in order to study the generation of dynamics of the tensor fields within the composite models \labelcref{eq:RewrittenFerminAction,eq:RewrittenScalarAction}.

%%%%%%%%%%%%%%%%%%%%%%%%%%%%%%%%%%%%%
\subsection{Fermionic model}
%%%%%%%%%%%%%%%%%%%%%%%%%%%%%%%%%%%%%

For the fermionic model \labelcref{eq:RewrittenFerminAction} we make the following ansatz for the effective action in Euclidean space:
\begin{align}
\label{eq:Feffectiveaction}
\Gamma^\psi_{k} &= \int\df^4x\Bigg[\frac{1}{2}\bar\psi \overleftrightarrow{\Slash \p} \psi
+ \kappa_\psi\, T^{(f)}_{\mu\nu} H_{\mu\nu}
- \lambda_2 H_{\mu\nu}H_{\mu\nu} 
- \lambda_0 H^2
\nn
&\quad
+\frac{Z_{H0}}{4} (\p_\lambda H_{\mu\nu})^2
+\frac{Z_{H1}}{4} \p_\lambda H_{\mu\nu}\p_\nu H_{\mu\lambda}\nn
&\quad
+\frac{Z_{H2}}{4} \p_\mu H_{\mu\nu}\p_\nu H
+ \frac{Z_{H3}}{4} (\p_\mu H)^2
\Bigg]\,.
\end{align}
Here we use the local potential approximation, fixing the field renormalization factor of the fermionic field to unity at all RG scales.
The initial condition at a certain UV scale $\Lambda$ for \labelcref{eq:Feffectiveaction} reads as $\kappa_\psi|_{\Lambda}=1$, $\lambda_2|_\Lambda=\lambda_2'$, $\lambda_0|_\Lambda=\lambda_0'$, and $Z_{Hi}|_\Lambda=0$ for all $i$. 

We neglect the RG running of $\kappa_\psi$ in the present work. Within our minimal truncation, $\kappa_\psi$ multiplies the operator $T^{(f)}_{\mu\nu}H_{\mu\nu}$ and does not introduce additional tensor structures. Since our main aim here is to demonstrate the dynamical generation of tensor-field kinetic terms from fermionic fluctuations, we keep $\kappa_\psi$ fixed in order to focus on this mechanism in its simplest setting.
We stress, however, that a fully coupled treatment including the flow of $\kappa_\psi$ would require evaluating its $\beta$ function and incorporating its feedback through the dynamical tensor propagator. This constitutes a nontrivial extension of the present truncation and will be addressed in future work.

Using the Wetterich equation \labelcref{eq:Wettericheq}, we obtain the flow equations for the couplings in the action \labelcref{eq:Feffectiveaction}.
At the truncation level of  \labelcref{eq:Feffectiveaction}, the loop corrections to the two-point function of $H_{\mu\nu}$ arise from the fermionic fluctuations.
The flow equation for the two-point function of $H_{\mu\nu}$ is depicted diagrammatically as 
\begin{align}
\label{eq:floweqTwopoint}
\p_t\left( \Gamma_k^{\psi,(2)}(p)\right)_{\mu\nu\rho\sigma}
&:=\frac{\delta^2 (\p_t\Gamma_k^\psi) }{\delta H_{\mu\nu}(p)\delta H_{\rho\sigma}(-p)}\nn[3ex]
&=-\,\vcenter{\hbox{\includegraphics[width=30mm]{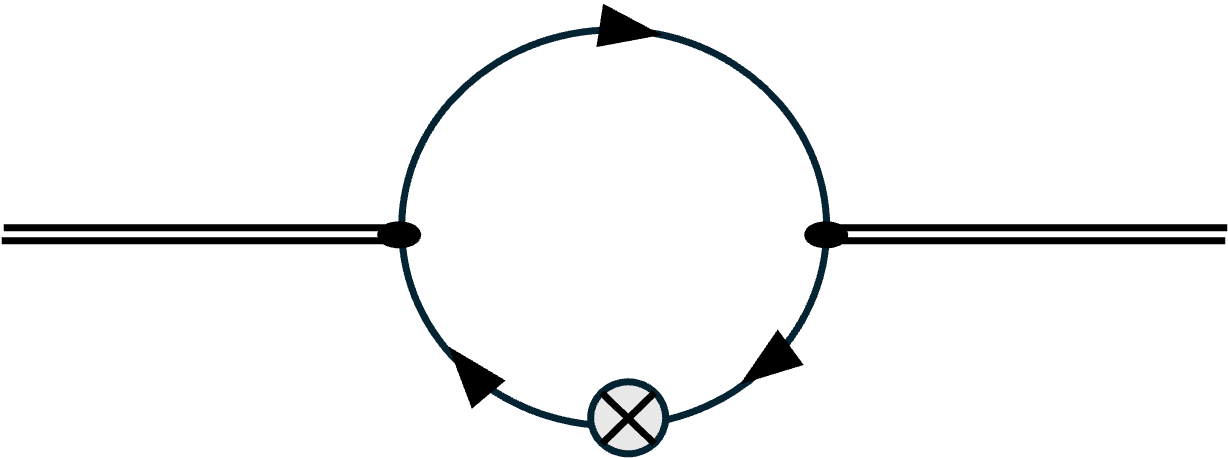}
\put(-65,25){\makebox(0,0)[r]{\strut{}$p\to$}}
\put(-0,25){\makebox(0,0)[r]{\strut{}$p\to$}}
\put(-32,38){\makebox(0,0)[r]{\strut{}$p+q$}}
\put(-30,-2){\makebox(0,0)[r]{\strut{}$q$}}
\put(-50,-2){\makebox(0,0)[r]{\strut{}$q$}}
}\vskip1mm}\,,
\end{align}
where $p$ is the external momentum; the single and double solid lines denote the fermionic fields $\psi$, $\bar\psi$, and tensor field $H_{\mu\nu}$, respectively; and the cross circle is the cutoff insertion $\p_t\mathcal R_k$.
The use of the Wetterich equation \labelcref{eq:Wettericheq} yields
\begin{align}
\label{eq:floweqTwopointexplicit}
&\p_t\left(\Gamma_k^{\psi,(2)}(p)\right)_{\mu\nu\rho\sigma} \nn
&= -2\kappa_\psi^2\int\frac{\df^4q}{(2\pi)^4} {\mathcal T}^{\psi}_{\mu\nu\rho\sigma}(q,p) f_k^\psi(q,p)\p_t r_k^\psi(|q|/k)\,. 
%&= {\mathcal F}^{(0)}_{\mu\nu\rho\sigma} +  {\mathcal F}^{(2)}_{\mu\nu\rho\sigma} p^2+\cdots 
\end{align}
Here $r_k^\psi$ is defined in \labelcref{eq:fermioncut}, and $f^\psi$ and ${\mathcal T}^{\psi}_{\mu\nu\rho\sigma}$ are given respectively by
\begin{align}
&f_k^\psi(q,p)= \frac{1}{q^4(1+r_k^\psi)^2 (p+q)^2(1+\tilde r_k^\psi)} \,,\\
& {\mathcal T}^{\psi}_{\mu\nu\rho\sigma}(q,p) 
= \frac{N_f}{16} \,\tr\Big[\Slash q \Big((2q+p)_{\mu} \gamma_{\nu} + (2q+p)_\nu \gamma_\mu \Big)\nn
&\qquad \times (\Slash p + \Slash q)\Big((2q+p)_\rho \gamma_\sigma + (2q+p)_\sigma \gamma_\rho \Big) \Slash q\Slash q\Big]\,,
\end{align}
with $\tilde r_k^\psi=r_k^\psi(|q+p|/k)$.
Performing the Taylor expansion in terms of $p$ and projecting out the tensor structure in \labelcref{eq:floweqTwopointexplicit}, we extract the flow equations for each coupling.
The left-hand side of \labelcref{eq:floweqTwopointexplicit} at $p=0$ is $\p_t\big( \Gamma_k^{\psi,(2)}(0)\big)_{\mu\nu\rho\sigma}= -\p_t \lambda_2 (\delta_{\mu\rho}\delta_{\nu\sigma} + \delta_{\mu\sigma}\delta_{\nu\rho}) - 2\p_t\lambda_0 \delta_{\mu\nu}\delta_{\rho\sigma}$.
The flow equations for the mass parameters are obtained from the $O(p^0)$ term and are given by
\begin{subequations}
\begin{align}
\p_t \lambda_0 &= k^4 \frac{1}{36}\frac{N_f\kappa_\psi^2}{(4\pi)^2}\,,\\[1ex]
\p_t \lambda_2 &= -k^4\frac{1}{36}\frac{N_f\kappa_\psi^2}{(4\pi)^2}\,.
\end{align}
\end{subequations}
Next, we show the flow equations for the field renormalization factors $Z_{Hi}$.
The left-hand side of \labelcref{eq:floweqTwopointexplicit} at order $O(p^2)$ is given by
\begin{align}
&\p_t\left(\Gamma_k^{\psi,(2)}(p)\right)_{\mu\nu\rho\sigma} \overset{O(p^2)}{=}
\frac{\p_t Z_{H0}}{2}\,p^2\,\delta_{\mu\rho}\,\delta_{\nu\sigma} \nn
&\quad
+\frac{\p_t Z_{H1}}{4}\Big(
%p_{\rho}p_{\nu}\,\delta_{\mu\sigma}+
p_{\sigma}p_{\nu}\,\delta_{\mu\rho}
+ p_{\rho}p_{\mu}\,\delta_{\nu\sigma}
%+ p_{\sigma}p_{\mu}\,\delta_{\nu\rho}
\Big) \nn
&\quad
+\frac{\p_t Z_{H2}}{4}\Big(
p_{\mu}p_{\nu}\,\delta_{\rho\sigma}
+ p_{\rho}p_{\sigma}\,\delta_{\mu\nu}
\Big)\nn
&\quad
+\frac{\p_t Z_{H3}}{2}\,p^2\,\delta_{\mu\nu}\,\delta_{\rho\sigma}\,.
\end{align}
After performing the momentum integral over $q$ in the right-hand side of \labelcref{eq:floweqTwopointexplicit} and identifying the corresponding tensor structures in $\mathcal T_{\mu\nu\rho\sigma}^\psi(q,p)$, we obtain the flow equations for the kinetic terms, i.e., the coefficients of the $O(p^2)$ contributions.
They take the following form:
\begin{align}
\p_t Z_{Hi} =k^2 A_i\frac{N_f\kappa_\psi^2}{(4\pi)^2}\,.
\end{align}
In the current setup, we obtain the factors $A_i$ as
\begin{align}
\label{eq:factors for fermion}
&A_{0}=-\frac{7}{48}\,,&
&A_{1}=\frac{1}{4}\,,&
&A_{2}=-\frac{5}{24}\,,&
&A_{3}=\frac{7}{96}\,.
\end{align}
%

%%%%%%%%%%%%%%%%%%%%%%%%%%%%%%%%%%%%%
\subsection{Scalar model}
%%%%%%%%%%%%%%%%%%%%%%%%%%%%%%%%%%%%%

We assume the effective action for the scalar model \labelcref{eq:RewrittenScalarAction} to be
\begin{align}
\label{eq:Beffectiveaction}
\Gamma^\phi_{k} &= \int\df^4x\Bigg[\frac{1}{2} \p_\mu \vec\phi \cdot \p_\mu \vec\phi
+ \kappa_\phi T^{(s)}_{\mu\nu} C_{\mu\nu}
- \xi_2 C_{\mu\nu}C_{\mu\nu} 
- \xi_0 C^2
\nn
&\quad
+\frac{Z_{C0}}{4} (\p_\lambda C_{\mu\nu})^2
+\frac{Z_{C1}}{4} \p_\lambda C_{\mu\nu}\p_\nu C_{\mu\lambda}
\nn
&\quad
+\frac{Z_{C2}}{4} \p_\mu C_{\mu\nu}\p_\nu C
+ \frac{Z_{C3}}{4} (\p_\mu C)^2
\Bigg]\,,
\end{align}
where we neglect the running effect of $\kappa_\phi$ and the field renormalization of the scalar field as the same reason as the fermionic case.

In the same manner as in the fermionic case, let us derive the flow equations for \labelcref{eq:Beffectiveaction}.
The flow equation for the two-point function of $C_{\mu\nu}$ is represented diagrammatically as
\begin{align}
\label{eq:floweqTwopointS}
\p_t \left(\Gamma^{\phi,(2)}_k(p)\right)_{\mu\nu\rho\sigma}
&:=\frac{\delta^2 (\p_t\Gamma_k^\phi) }{\delta C_{\mu\nu}(p)\delta C_{\rho\sigma}(-p)}\nn[3ex]
&= \frac{1}{2}\,\,\vcenter{\hbox{\includegraphics[width=30mm]{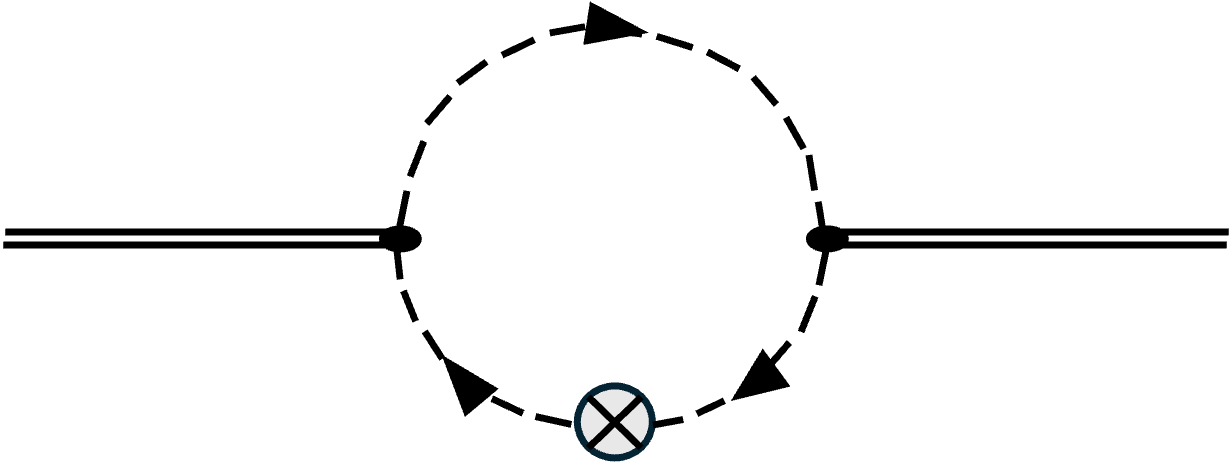}
\put(-65,25){\makebox(0,0)[r]{\strut{}$p\to$}}
\put(-0,25){\makebox(0,0)[r]{\strut{}$p\to$}}
\put(-32,38){\makebox(0,0)[r]{\strut{}$p+q$}}
\put(-30,-2){\makebox(0,0)[r]{\strut{}$q$}}
\put(-50,-2){\makebox(0,0)[r]{\strut{}$q$}}
}\vskip1mm}\,,
\end{align}
where the dashed and double solid lines denote the scalar field $\vec\phi$ and the tensor field $C_{\mu\nu}$, respectively.
The flow equation for the scalar two-point function reads as
\begin{align}
&\p_t \left(\Gamma^{\phi,(2)}_k(p)\right)_{\mu\nu\rho\sigma}\nn
&=\kappa_\phi^2 \int\frac{\df^4q}{(2\pi)^4}  {\mathcal T}^{\phi}_{\mu\nu\rho\sigma}(q,p)f_k^\phi(q,p) \p_t r_k^\phi(|q|/k)\,,
%&= {\mathcal S}^{(0)}_{\mu\nu\rho\sigma} +  {\mathcal S}^{(2)}_{\mu\nu\rho\sigma} p^2+\cdots  \,,
\end{align}
where $r_k^\phi$ is given in \labelcref{eq:scalarcut} and we have defined 
\begin{align}
&f_k^\phi(q,p) = \frac{1}{q^4(1+r^\phi_k)^4(q+p)^2(1+\tilde r_k^\phi)^2} \,,\\
&{\mathcal T}^{\phi}_{\mu\nu\rho\sigma}(q,p)= N\Big[q_\mu (q+p)_\nu + q_\nu (q+p)_\mu -\delta_{\mu\nu} q\cdot (q+p)\Big] \nn
&\quad \times \Big[q_\rho (q+p)_\sigma + q_\sigma (q+p)_\rho -\delta_{\rho\sigma} q\cdot (q+p)\Big] q^2\,,
\end{align}
with $\tilde r_k^\phi=r_k^\phi(|p+q|/k)$.
From the $O(p^0)$ term, the flow equation governing the two-point function of $C_{\mu\nu}$ yields the corresponding flow of the mass parameters as 
\begin{subequations}
\begin{align}
\p_t \xi_2 &=- k^4 \frac{1}{6}\frac{N\kappa_\phi^2}{(4\pi)^2}\,,\\[1ex]
\p_t \xi_0 &= -k^4\frac{1}{12}\frac{N\kappa_\phi^2}{(4\pi)^2}\,,
\end{align}
\end{subequations}
while from the $O(p^2)$ term, we obtain
\begin{align}
\p_t Z_{Ci} = k^2B_i\frac{N\kappa_\phi^2}{(4\pi)^2}\,.
\end{align}
Here the factors are found to be
\begin{align}
\label{eq:factors for scalar}
&B_{0}=-\frac{1}{3}\,,&
&B_{1}=\frac{4}{3}\,,&
&B_{2}=-\frac{2}{3}\,,&
&B_{3}=-\frac{1}{6}\,.
\end{align}

We close this section with a remark.
Regarding the role of equations of motion, we only employ the classical equation of motion for the auxiliary field $H_{\mu\nu}$ (or $C_{\mu\nu}$) to invert the Hubbard-Stratonovich transformation, i.e., to demonstrate explicitly that integrating out the auxiliary field reproduces the original $(T_{\mu\nu})^2$ interaction. This should not be confused with on shell conditions for the fundamental fields.

The Hubbard-Stratonovich transformation is an exact rewriting at the level of the functional integral and does not modify the global space-time translation symmetry of the theory. Accordingly, the conserved Noether current associated with translations is unchanged (up to terms that vanish upon using the auxiliary-field equation of motion) when all fields, including the auxiliary ones, are taken into account.

The fRG flow yields $\Gamma_k[\psi,H_{\mu\nu}]$ (or $\Gamma_k[\phi,C_{\mu\nu}]$) off shell, so that we do not use classical on shell relations in the course of the flow. The energy-momentum tensor entering our truncation should therefore be understood as defining the composite channel in theory space, rather than as the exact Noether current of the truncated effective action. Quantum equations of motion arise only from $\Gamma_k$ at $k\to 0$, and are relevant for determining the physical background rather than for the derivation of the flow.

%%%%%%%%%%%%%%%%%%%%%%%%%%%%%%%%%%%%
\section{Analysis}
\label{sec:Analysis}
%%%%%%%%%%%%%%%%%%%%%%%%%%%%%%%%%%%%

Solving the flow equations for $\lambda_2$, $\lambda_0$, and $Z_{Hi}$ with the initial condition $\lambda_2|_{k=\Lambda}=\lambda_2'$, $\lambda_0|_{k=\Lambda}=\lambda_0'$, and $Z_{Hi}|_{k=\Lambda}=0$ in the fermionic case, we obtain
\begin{align}
\lambda_2(k) &= \lambda_2' +\frac{N_f}{4(4\pi)^2} \frac{4}{9}(k^4-\Lambda^4)\,,\\
\lambda_0(k)&= \lambda_0'  +\frac{N_f}{4(4\pi)^2} \frac{-4}{9}(k^4-\Lambda^4)\,,\\
Z_{Hi}(k)&= \frac{N_fA_i}{2(4\pi)^2}(k^2-\Lambda^2)\,.
\end{align}
In the same manner, in the scalar case we have
\begin{align}
\xi_2(k) &= \xi_2' +\frac{N}{4(4\pi)^2} \frac{-2}{3}(k^4-\Lambda^4)\, ,\\
\xi_0(k)&= \xi_0' +\frac{N}{4(4\pi)^2} \frac{-1}{3}(k^4-\Lambda^4)\, ,\\
Z_{Ci}(k)&= \frac{NB_i}{2(4\pi)^2}(k^2-\Lambda^2)\,,
\end{align}
where the initial condition was set so as to be $\xi_2|_{k=\Lambda}=\xi_2'$, $\xi_0|_{k=\Lambda}=\xi_0'$ and $Z_{Ci}|_{k=\Lambda}=0$.

First, we tune the bare mass parameters such that the tensor fields become massless in the IR limit ($k=0$). 
This amounts to subtracting the power-divergent contributions generated along the flow.
Thus, the bare mass parameters are chosen such that the loop corrections cancel.
In the fermionic case, we give
\begin{align}
\lambda_2' &= \frac{N_f}{4(4\pi)^2} \frac{4}{9}\Lambda^4\,,&
\lambda_0' &= -\frac{N_f}{4(4\pi)^2} \frac{4}{9}\Lambda^4\,,
\end{align}
while in the scalar case, we give
\begin{align}
\xi_2' &=\frac{N}{4(4\pi)^2} \frac{-2}{3}\Lambda^4\,, &
\xi_0' &=\frac{N}{4(4\pi)^2} \frac{-1}{3}\Lambda^4\,.
\end{align}
These subtractions correspond to the additive renormalization of the power divergences.

Next, we consider the kinetic terms of the tensor fields in the IR regime.
To this end, we renormalize the fields as $\sqrt{Z_{H0}(0)}H_{\mu\nu}=h_{\mu\nu}$ and $\sqrt{Z_{C0}(0)}C_{\mu\nu}=c_{\mu\nu}$ in the fermionic and scalar cases, respectively.
Then, the kinetic terms of $h_{\mu\nu}$ and $c_{\mu\nu}$ are given respectively by
\begin{align}
\label{eq:fermionkine}
\mathcal L_h &=\frac{1}{4} (\p_\lambda h_{\mu\nu})^2
+\frac{Z_{H1}(0)}{4Z_{H0}(0)} \p_\lambda h_{\mu\nu}\p_\nu h_{\mu\lambda}\nn
&\quad
+\frac{Z_{H2}(0)}{4Z_{H0}(0)} \p_\mu h_{\mu\nu}\p_\nu h
+ \frac{Z_{H3}(0)}{4Z_{H0}(0)} (\p_\mu h)^2,
\\[2ex]
\label{eq:scalarkine}
\mathcal L_c &=\frac{1}{4} (\p_\lambda c_{\mu\nu})^2
+\frac{Z_{C1}(0)}{4Z_{C0}(0)} \p_\lambda c_{\mu\nu}\p_\nu c_{\mu\lambda}
\nn
&\quad
+\frac{Z_{C2}(0)}{4Z_{C0}(0)} \p_\mu c_{\mu\nu}\p_\nu c
+ \frac{Z_{C3}(0)}{4Z_{C0}(0)} (\p_\mu c)^2\,.
\end{align}
Note that this field renormalization does not change the massless condition for the tensor fields, since the operation is multiplicative rather than additive.
Within these kinetic terms, the explicit dependence on the cutoff scale $\Lambda$, as well as common factors such as the number of the degrees of freedom ($N_f$ and $N$) and $(4\pi)^{-2}$, cancel out, leaving the coefficients dependent only on the ratios given in \labelcref{eq:factors for fermion} and \labelcref{eq:factors for scalar}.
More specifically, in the fermionic case, we have
\begin{subequations}
\label{eq:fermionCoe}
\begin{align}
&\frac{Z_{H1}(0)}{Z_{H0}(0)} = \frac{A_1}{A_0} = -\frac{12}{7}\,,\\[1ex]
&\frac{Z_{H2}(0)}{Z_{H0}(0)} = \frac{A_2}{A_0} =\frac{10}{7}\,,\\[1ex]
&\frac{Z_{H3}(0)}{Z_{H0}(0)} = \frac{A_3}{A_0} =-\frac{1}{2}\,,
\end{align}
\end{subequations}
while in the scalar case, we have
\begin{subequations}
\label{eq:scalarCoe}
\begin{align}
&\frac{Z_{C1}(0)}{Z_{C0}(0)} = \frac{B_1}{B_0} =-4\,,\\[1ex]
&\frac{Z_{C2}(0)}{Z_{C0}(0)} = \frac{B_2}{B_0} =2\,,\\[1ex]
&\frac{Z_{C3}(0)}{Z_{C0}(0)} = \frac{B_3}{B_0} =\frac{1}{2}\,.
\end{align}
\end{subequations}

Here, the diffeomorphism invariant kinetic terms of a metric field $g_{\mu\nu}$ around a flat background $\delta_{\mu\nu}$ read  
\begin{align}
\label{eq:EHPropagator}
\mathcal L_\text{diff}&= \frac{1}{4} (\p_\lambda g_{\mu\nu})^2
-\frac{1}{2} \p_\lambda g_{\mu\nu}\p_\nu g_{\mu\lambda}\nn
&\quad
+\frac{1}{2} \p_\mu g_{\mu\nu}\p_\nu g
-\frac{1}{4} (\p_\mu g)^2\,,
\end{align}
where $g=g_{\mu\nu}\delta_{\mu\nu}$.
Obviously, both \cref{eq:fermionkine} with \labelcref{eq:fermionCoe} and \cref{eq:scalarkine} with \labelcref{eq:scalarCoe} differ from the diffeomorphism invariant kinetic term \labelcref{eq:EHPropagator}.
This suggests that the induced kinetic terms can be interpreted as gauge-fixed forms of a diffeomorphism-invariant kinetic operator. 
However, no choice of the covariant gauge-fixing parameters $\alpha$ and $\beta$ reproduces the coefficients obtained here.
To see this, let us employ the covariant gauge fixing condition
\begin{align}
\label{eq:gaugefixed}
\mathcal L_\text{gf} = \frac{1}{2\alpha}\left(\p_\mu g_{\mu\nu} -\frac{1+\beta}{4} \p_\nu g \right)^2\,,
\end{align}
where $\alpha$ and $\beta$ are gauge-fixing parameters.
Note that the choice $\beta=1$ corresponds to the so-called de Donder (or harmonic) gauge.
Together with the gauge fixing term \labelcref{eq:gaugefixed}, \cref{eq:EHPropagator} reads as
\begin{align}
\label{eq:gaugefixedLag}
\mathcal L_\text{diff} + \mathcal L_\text{gf} 
&=\frac{1}{4} (\partial_\lambda g_{\mu\nu})^2
- \frac{1}{2}\left( 1-\frac{1}{\alpha} \right)\,\partial_\lambda g_{\mu\nu}\,\partial_\nu g_{\mu\lambda} \nn
&\quad
+ \frac{1}{2}\left(1 - \frac{1+\beta}{2\alpha}\right)\,
\partial_\mu g_{\mu\nu}\,\partial_\nu g \nn
&\quad
-\frac{1}{4} \left(1 - \frac{(1+\beta)^2}{8\alpha}\right)\,
(\partial_\mu g)^2\,.
\end{align}
However, comparing \labelcref{eq:gaugefixedLag} with \labelcref{eq:fermionkine} and \labelcref{eq:scalarkine} one finds no appropriate choice of $\alpha$ and $\beta$ matching with the induced kinetic terms in both fermionic and scalar cases.

Another possible gauge-fixing term may be given by
\begin{align}
\label{eq:anothergaugefixing}
\mathcal L_\text{gf} = \frac{1}{2\zeta_1} \partial_\lambda g_{\mu\nu}\,\partial_\nu g_{\mu\lambda} - \frac{1}{\zeta_2}\partial_\mu g_{\mu\nu}\,\partial_\nu g + \frac{1}{2\zeta_3}(\partial_\mu g)^2\,,
\end{align}
with three gauge fixing parameters $\zeta_1$, $\zeta_2$ and $\zeta_3$ although it is not covariant.
In this case, the induced kinetic terms can be formally matched to \eqref{eq:EHPropagator} supplemented by \eqref{eq:anothergaugefixing} for specific choices of $\zeta_i$.
More specifically, the kinetic terms \labelcref{eq:fermionkine} and \labelcref{eq:scalarkine} correspond to \labelcref{eq:EHPropagator} plus \labelcref{eq:anothergaugefixing} with $\zeta_1=7$, $\zeta_2=7$ and $\zeta_3=4$ for the fermionic case, while $\zeta_1=-1$, $\zeta_2=\infty$ and $\zeta_3=4/3$ for the scalar case.

We find that agreement with the diffeomorphism-invariant quadratic Einstein-Hilbert structure can, at present, be established only in the transverse-traceless spin-2 sector.
The remaining contributions reside in the longitudinal and trace sectors.
Although these terms can be cast into a form reminiscent of structures that arise in gauge-fixed presentations of the quadratic
Einstein-Hilbert action, we do not assign to them the interpretation of a true gauge-fixing contribution in the present framework.

\begin{figure}
    \centering
    \includegraphics[width=0.8\linewidth]{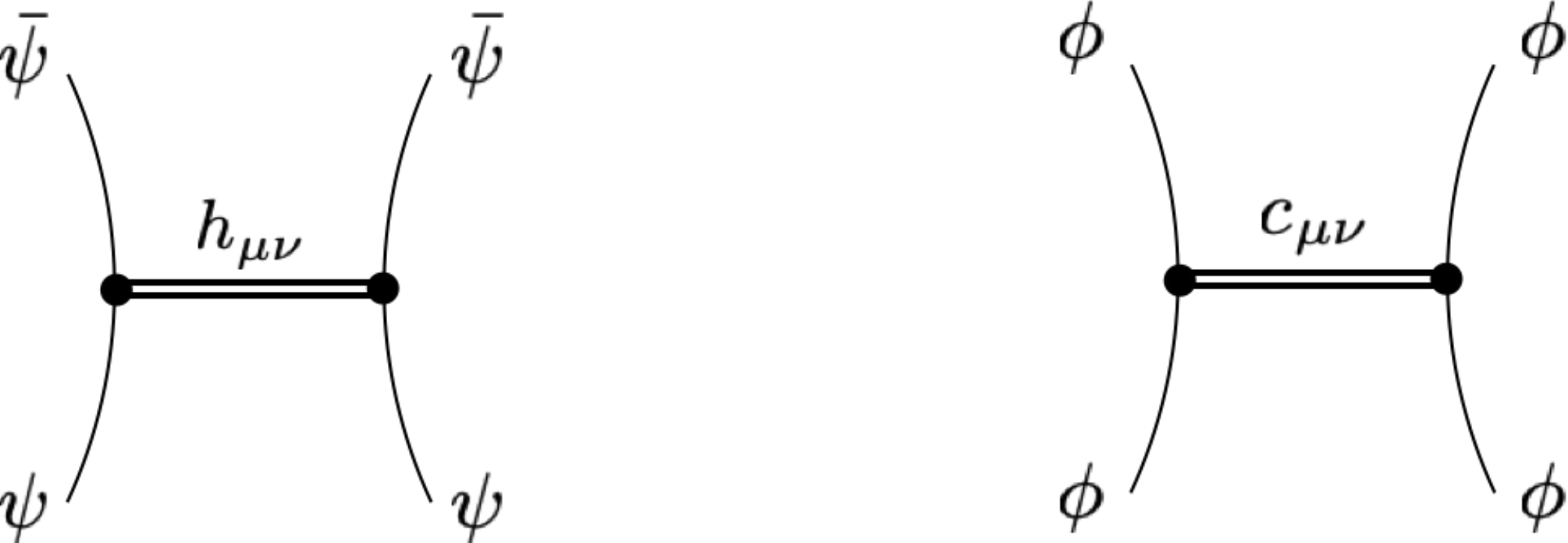}  
    \caption{Single graviton exchange process in the fermionic (left) and scalar (right) theories.}
    \label{fig:exchange}
\end{figure}

We close this section with remarks on the gravitational interaction with a matter field. 
After renormalizing the tensor fields, the couplings between the metric fluctuations and the energy-momentum tensor in \labelcref{eq:Feffectiveaction} and \labelcref{eq:Beffectiveaction} take the form
\begin{align}
\mathcal L^{Th} &= \frac{\kappa_\psi}{\sqrt{Z_{H0}(0)}} T^{(f)}_{\mu\nu} h_{\mu\nu}\,,\\
\mathcal L^{Tc} &= \frac{\kappa_\phi}{\sqrt{Z_{C0}(0)}}T^{(s)}_{\mu\nu} c_{\mu\nu}\,.
\end{align}
From these expressions, one may define an effective gravitational coupling by matching the single-graviton exchange amplitude 
(see Fig.~\ref{fig:exchange}). 
Within this definition, we obtain $G_N=\kappa_\psi^2/(4\pi Z_{H0}(0))$ for the fermionic case and $G_N=\kappa_\phi^2/(4\pi Z_{C0}(0))$ for the scalar case.

%%%%%%%%%%%%%%%%%%%%%%%%%%%%%%%%%%
%%%%%%%%%%%%%%%%%%%%%%%%%%%%%%%%%%

%%%%%%%%%%%%%%%%%%%%%%%%%%%
\section{Summary}
\label{sec:summary}
%%%%%%%%%%%%%%%%%%%%%%%%%%%

In this work, we have taken a first step toward constructing a gravitational framework in which the graviton emerges as a composite field within the fRG approach. 
We analyzed fermionic and scalar matter models separately and derived the flow equations governing the kinetic terms of auxiliary tensor fields associated with composite operators. 
Although these fields are nondynamical at tree level, quantum fluctuations of the matter sector generate finite kinetic terms in the infrared.

The induced kinetic structure agrees with the diffeomorphism-invariant quadratic Einstein-Hilbert form in the transverse-traceless spin-2 sector. 
The remaining contributions belong to the longitudinal and trace sectors; while they can be rewritten in a form reminiscent of gauge-fixed treatments of the quadratic Einstein–Hilbert action, we refrain from interpreting them as genuine gauge-fixing terms within the present truncation.

A complete understanding of emergent gravitational dynamics in composite models requires the computation of higher-order interaction vertices, such as three-graviton and matter-graviton vertices. 
The fRG framework provides a suitable tool for this purpose. 
In particular, dynamical bosonization within the fRG offers a promising strategy for systematically accessing these vertices~\cite{Gies:2001nw, Gies:2002hq, Pawlowski:2005xe, Floerchinger:2009uf, Braun:2014ata, Mitter:2014wpa, Cyrol:2017qkl, Cyrol:2017ewj, Alkofer:2018guy, Denz:2019ogb, Fu:2019hdw, Goertz:2024dnz}. 
A detailed investigation of this program is left for future work.

\bibliographystyle{JHEP}
\bibliography{refs}
\end{document}